\documentclass[fleqn,10pt]{wlscirep}
\usepackage{amsmath}
\usepackage{soul}

\title{Carbon nanotube synthesis and spinning as macroscopic fibers assisted by the ceramic reactor tube}

\author[1,\dag]{X. Rodiles}
\author[1,\dag]{V. Reguero}
\author[1]{M. Vila}
\author[1]{B. Alemán}
\author[1]{L. Arévalo}
\author[2]{F. Fresno}
\author[2,*]{V.A. de la Peña  O'Shea}
\author[1,*]{J.J. Vilatela}
\affil[1]{IMDEA Materials Institute, 28906 Getafe, Madrid, Spain}
\affil[2]{IMDEA Energy Institute, 28935 Móstoles, Madrid, Spain}
\affil[$\dag$] {These authors contributed equally to this work}
\affil[*]{Corresponding author email: victor.delapenya@imdea.org, juanjose.vilatela@imdea.org}

\begin{abstract}
Macroscopic fibers of carbon nanotubes (CNT) have emerged as an ideal architecture to exploit the exceptional properties of CNT building blocks in applications ranging from energy storage to reinforcement in structural composites. Controlled synthesis and scalability are amongst the most pressing challenges to further materialize the potential of CNT fibers. This work shows that under floating catalyst chemical vapor conditions in the direct spinning method, used both in research and industry, the ceramic reactor tube plays an unsuspected active role in CNT growth, leading for example to doubling of reaction yield when mullite (Al\textsubscript{4+2x}Si\textsubscript{2-2x}O\textsubscript{10-x}(x $\approx 0.4)$) is used instead of alumina (Al\textsubscript{2}O\textsubscript{3}), but without affecting CNT morphology in terms of number of layers, purity or degree of graphitization. This behaviour has been confirmed for different carbon sources and when growing either predominantly single-walled or multi-walled CNTs by adjusting promotor concentration. Analysis of large Si-based impurities occasionally found in CNT fiber fabric samples, attributed to reactor tube fragments that end up trapped in the porous fibers, indicate that the role of the reactor tube is in catalyzing the thermal decomposition of hydrocarbons, which subsequently react with floating Fe catalyst nanoparticles and produce extrusion of the CNTs and formation of an aerogel. Reactor gas analysis confirms that extensive thermal decomposition of the carbon source occurs in the absence of Fe catalyst particles, and that the concentration of different carbon species (e.g. carbon dioxide and ethylene) is sensitive to the reactor tube type. These finding open new avenues for controlled synthesis of CNT fibers by decoupling precursor decomposition from CNT extrusion at the catalyst particle.
\end{abstract}
\begin{document}

\flushbottom
\maketitle
%
%
\thispagestyle{empty}

\section*{Introduction}
Carbon nanotubes (CNTs) are extraordinary building blocks due to their exceptional combination of mechanical, electrical and thermal properties along the tube axis. They can be produced in large quantities as a highly graphitic material with a well defined structure, in some cases with molecular composition control in terms of number of layers, diameter of nanotubes and chiral angle. Their assembly into fibers and further arrays (yarns, tows, fabrics) has unlocked numerous applications in diverse fields, including electrodes for energy storage \cite{senokos2016}, neural stimulation \cite{Neural} and water desalination \cite{Cleis}, composite reinforcement elements \cite{Gonzalez2017}, sensors  \cite{Toribio2016}, and multiple wearable devices \cite{ReviewQingwen}.

This has been possible through the development of fiber spinning processes that can reproducibly produce uniform highly controlled materials. An example is the direct spinning method, whereby an aerogel of CNTs is produced by floating catalyst chemical vapor deposition (FC-CVD) and directly drawn out of a reactor and wound continuously as a macroscopic fiber \cite{Windle2004}. It is of particular interest because it combines independent control of CNT alignment,\cite{Aleman2015} composition in terms of number of layers \cite{Reguero2014} and CNT morphology, \cite{Selenio2016} with large scale production in laboratory setting ($>10$ km/day) and at industrial scaled-up facilities. CNT fibers are stronger than carbon fibers \cite{Science}, with higher thermal conductivity than copper \cite{copper12017,Science2013} and higher mass-normalized electrical conductivity than most metals \cite{Science2013,Lewaka2014} have been produced using this method.

There have been substantial efforts to understand the key parameters during synthesis and spinning that control CNT fiber structure and resulting properties. The process shares some features of conventional substrate-based CVD growth of CNT, both being based on catalytic decomposition of a carbon source and extrusion as a CNT from a transition metal catalyst particle, typically in a reducing (H\textsubscript{2}) atmosphere.

In the case of traditional substrate growth of CNTs by CVD, control over CNT type, quality and yield depend strongly  on  the presence and choice of a MO\textsubscript{x} support layer. The MO\textsubscript{x} acts as a barrier for catalyst diffusion but has also influence on the catalyst particle size/shape and catalytic activity \cite{Cantoro,Bronikowski2007}. The activity of supported catalyst vary depending on reducibility of the metal oxide and formation of silicides \cite{Wagner2007}. In the context of CNT growth from pre-deposied Fe catalyst particles on Si wafers with either SiO\textsubscript{2} or Al\textsubscript{2}O\textsubscript{3} thin coatings, for example, \textit{in-situ} studies have shown that CNT morphology and areal density strongly depend on  strength (i.e. charge transfer at) of the molten Fe nanoparticle/MOx interface \cite{Hofmann1,Hofman2}. In general, the use of a thin (100-400 nm) layer of SiO\textsubscript{2} has been an effective route to achieve reproducible synthesis of highly graphitic aligned, long CNTs on substrates ranging from Si wafers to carbon fibers \cite{Vilatela2015,Hermite2012}. 

On the other hand, FC-CVD is conducted at higher temperature ($>1000$ ºC) under conditions leading to the formation of catalyst nanoparticles "floating" in the gas phase, and which are assumed to entirely produce the decomposition of the carbon source and formation of carbon nanotubes\cite{moisala2003,moisala2005,moisala2006}. In the case of the direct spinning process, the presence of sulfur or another promoter enables the formation of an aerogel that can be continuously drawn out of the reactor during CNT growth to form a continuous macroscopic fiber. The promoter reduces carbon solubility in the catalyst, while also assists stabilization of the nascent CNT nucleus through the minimization of the interfacial energy between the CNT edge and the catalyst particle. CNT number of layers and yield are hence dependent on sulfur/carbon ratio (S/C) in the reaction, \cite{Reguero2014} thus indirectly also on the choice of sulfur precursor \cite{Sundaram2011}. These roles are similar to the effect of S and other group 16 elements as a “surfactant” in cast iron and graphitic poison in methanation reactions. In contrast, the concentration of Fe catalyst particles in the direct spinning process has a minor effect on CNT morphology or yield, provided there are sufficient to carry out the reaction, with only $< 0.01\%$ being active.

A prevalent assumption in CNT growth by FC-CVD is that the ceramic reactor tube, typically alumina, mullite or silica, plays no active role in the reaction and simply act as reaction vessels. However, here we show that the ceramic tube that hosts the FC-CVD plays an unsuspected active role in the FC-CVD reaction through carbon precursor decomposition. While it does not change CNT morphology, which remains controlled by the S/C ratio, it can double reaction yield when mullite (Al\textsubscript{4+2x}Si\textsubscript{2-2x}O\textsubscript{10-x}(x $\approx 0.4)$) is used instead of alumina (Al\textsubscript{2}O\textsubscript{3}), for example. Analysis of Si-based impurities, occasionally found in CNT fiber samples and presumed to detach from the reactor tube, helps understand general aspects of the reaction mechanism.

\section*{Results}
\subsection*{Synthesis in different reactor tubes}
\begin{figure}[ht]
\centering
\includegraphics[width=0.9\linewidth]{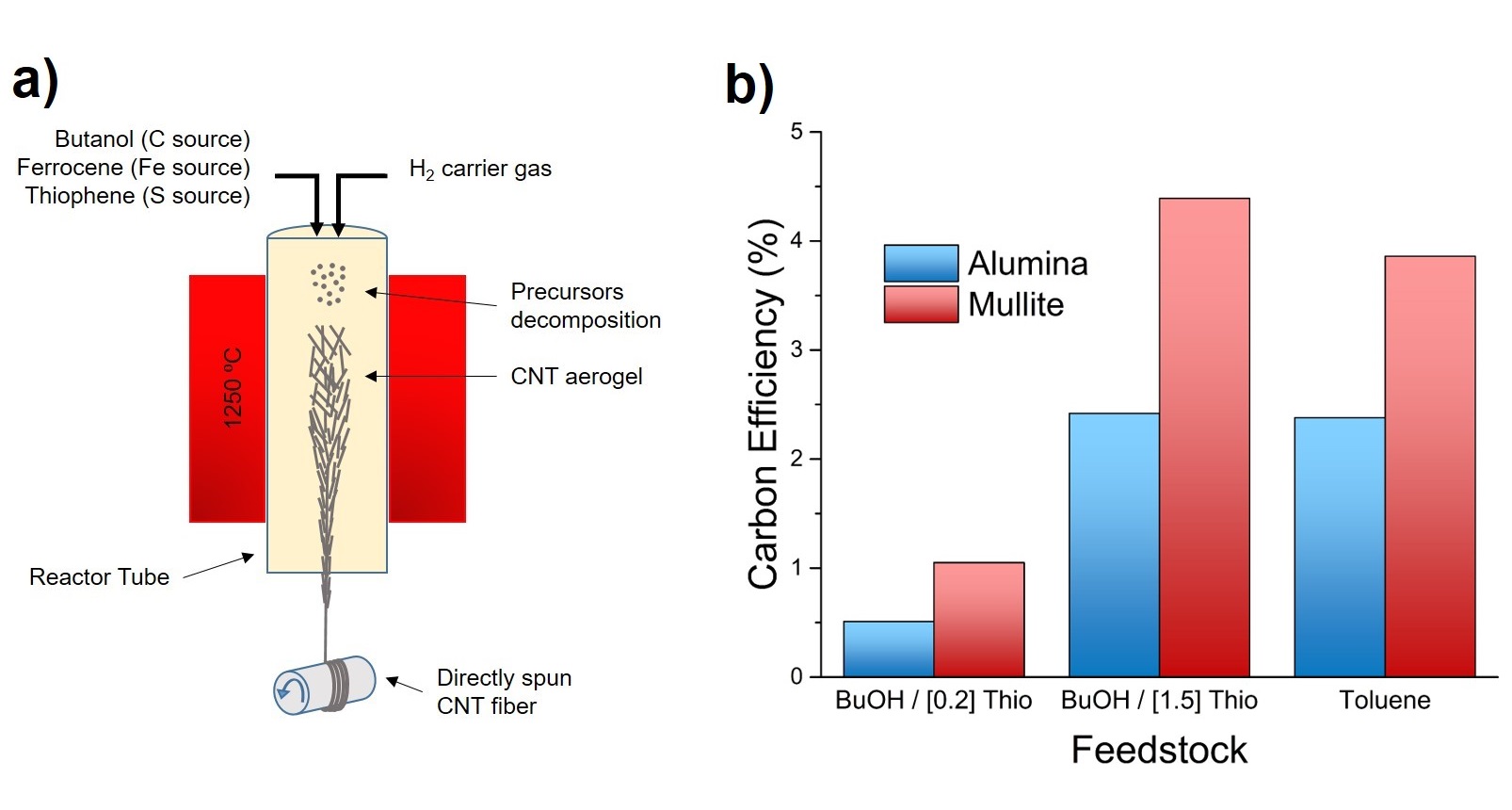}
\caption{Effects of reactor tube composition for the spinning CNT fibers directly from the gas phase during FC-CVD. a) Schematic of the reactor set-up. b) Carbon efficiency for mullite and allumina under different synthesis conditions including low (high) S/C ratio to produce SWNTs (MWNTs) or when using different carbon sources (butanol and toluene). Ferrocene used as Fe catalyst source}
\label{fig1}
\end{figure}

We have produced continuous CNT fibers in reactor tubes with different contents of SiO$_{2}$. In our typical set-up mullite (62.6$\%$ Al$_{2}$O$_{3}$, 35.2$\%$ SiO$_{2}$, 2.2$\%$ others) is used as a reactor tube, leading to the formation of a dark aerogel of CNTs that can be continuously spun for hours (Figure \ref{fig1}a). Interestingly, the use of an alumina reactor under otherwise identical synthesis (with ferrocence as catalyst source, thiophene as S promoter source and a liquid carbon source) conditions, while still leading to stable and continuous fiber spinning, produces a clear reduction in reaction conversion and changes in aerogel morphology.

This behaviour is observed under a wide range of synthesis conditions, for example when varying the S/C ratio so as to control CNT number of layers, or when using different carbon sources. Figure \ref{fig1}b shows consistently higher carbon efficiency in terms of conversion of injected carbon to graphitic carbon for mullite, under synthesis conditions to produce: predominantly single-walled carbon nanotubes (SWCNT), multi-walled carbon nanotubes (MWCNTs), and when using butanol or toluene as carbon sources.

\begin{figure}[ht]
\centering
\includegraphics[width=0.8\linewidth]{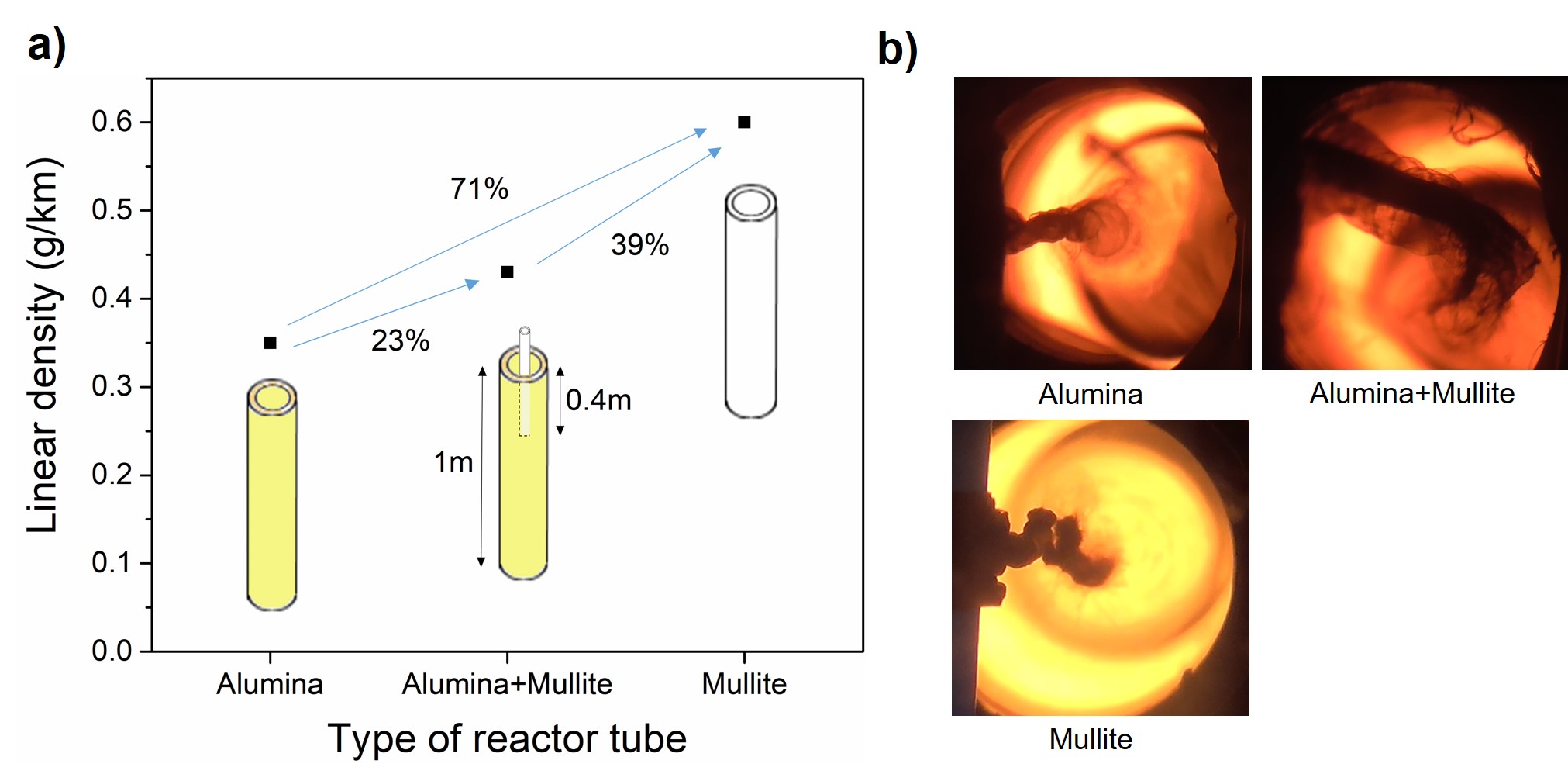}
\caption{Effects of reactor tube material on CNT aerogel properties. (a) Relationship between linear density (g/km) and the type of reactor tube. (b) Optical images showing increasing perceived thickness of the CNT aerogel when comparing reactions in alumina, alumina + mullite or mullite.}
\label{fig2}
\end{figure}

To investigate intermediate SiO$_{2}$ compositions we performed an additional experiment in which a mullite inner tube was gradually introduced from the top of the reactor into the alumina reactor tube, until the bottom of the inner mullite tube was in the region of $1250 ^{\circ}C$, as schematically shown in Figure \ref{fig2}a. The resulting CNT fiber had an intermediate linear density between those produced in alumina or mullite (Figure \ref{fig2}(a)).  At a spinning rate of 5 $m/min$, the linear mass density of the fiber produced in mullite is around $0.60 g/km$, higher than the fiber produced in alumina, which is $0.35 g/km$, whereas the fiber produced in alumina after insertion of a mullite tube had a linear density of $0.43 g/km$ (Figure \ref{fig2}(a)). These differences in linear density correlated well with the perceived darkness of the CNT aerogel observed from the bottom of the reactor, as shown in the photographs in Figure \ref{fig2}(b). No other major differences in the position or shape of the aerogel that could suggest growth of the CNTs on the ceramic surface, were observed. 

\begin{figure}[ht]
\centering
\includegraphics[width=0.5\linewidth]{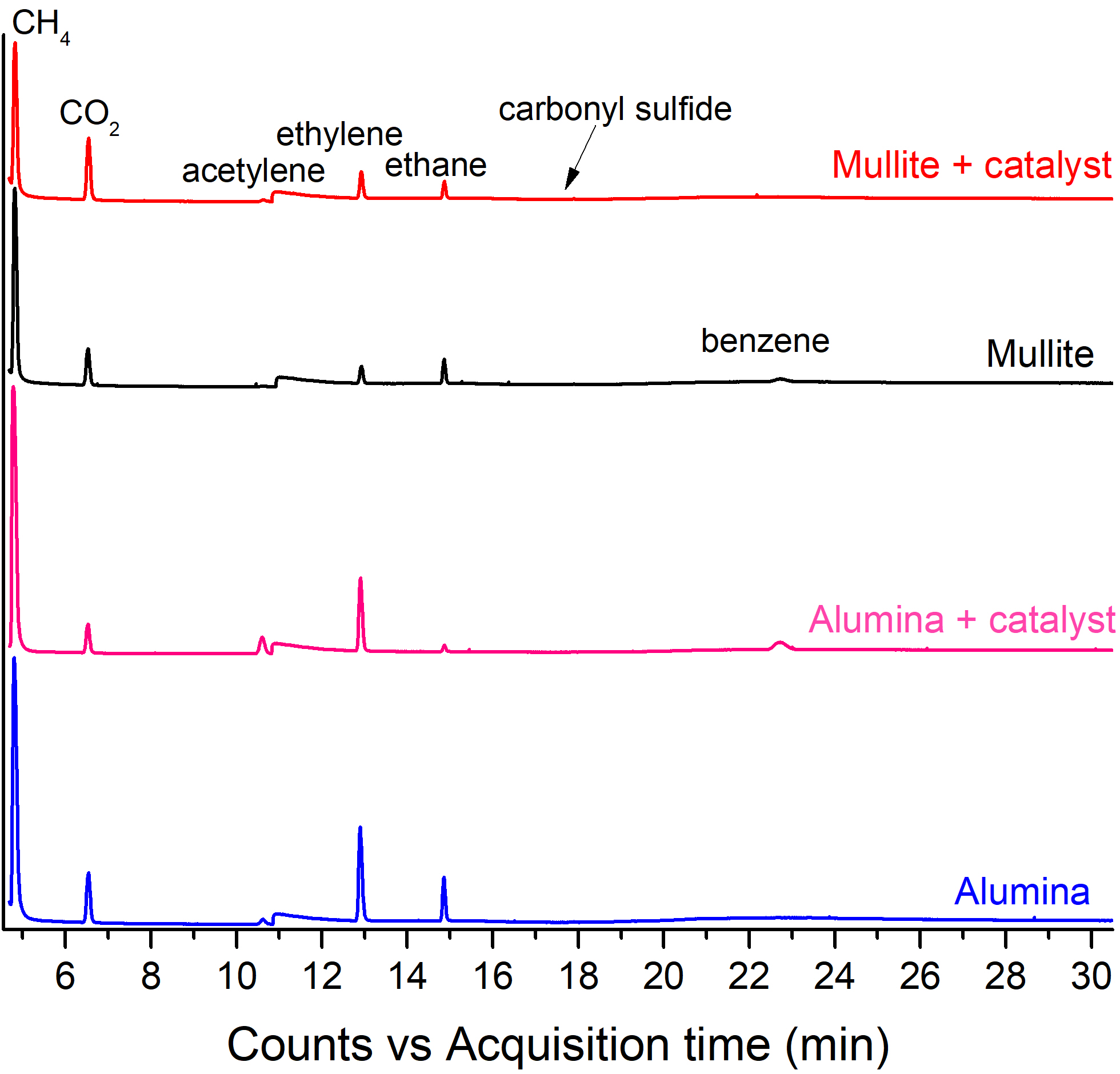}
\caption{Chromatographs of gases present at the outlet of the reactor. Four gases have been sampled in four different scenarios: when using either alumina or mullite tubes, and with or without catalyst. Chromatographs have been tiled vertically for clarity.}
\label{fig3_chrom}
\end{figure}

In order to shed light into the effect of the reactor tube on the reaction conditions, we studied the chemical composition of gases at the reactor outlet by performing chromatography measurements on gas samples extracted from the reactor when using mullite or alumina tubes. Reference measurements also included gas samples produced in the absence of Fe catalyst but with the same injection of the remaining precursors. At present, the experimental protocol used (see experimental section) only enables gas extraction from the lower part of the reactor, which can contain substantial nitrogen gas purposely introduced to replace hydrogen at the reactor exit for safety reasons, but which does not take part in the reaction. As a consequence, only relative concentrations of different species can be analyzed based on the chromatography data (Figure \ref{fig3_chrom}).

\begin{figure}[ht]
\centering
\includegraphics[width=0.8\linewidth]{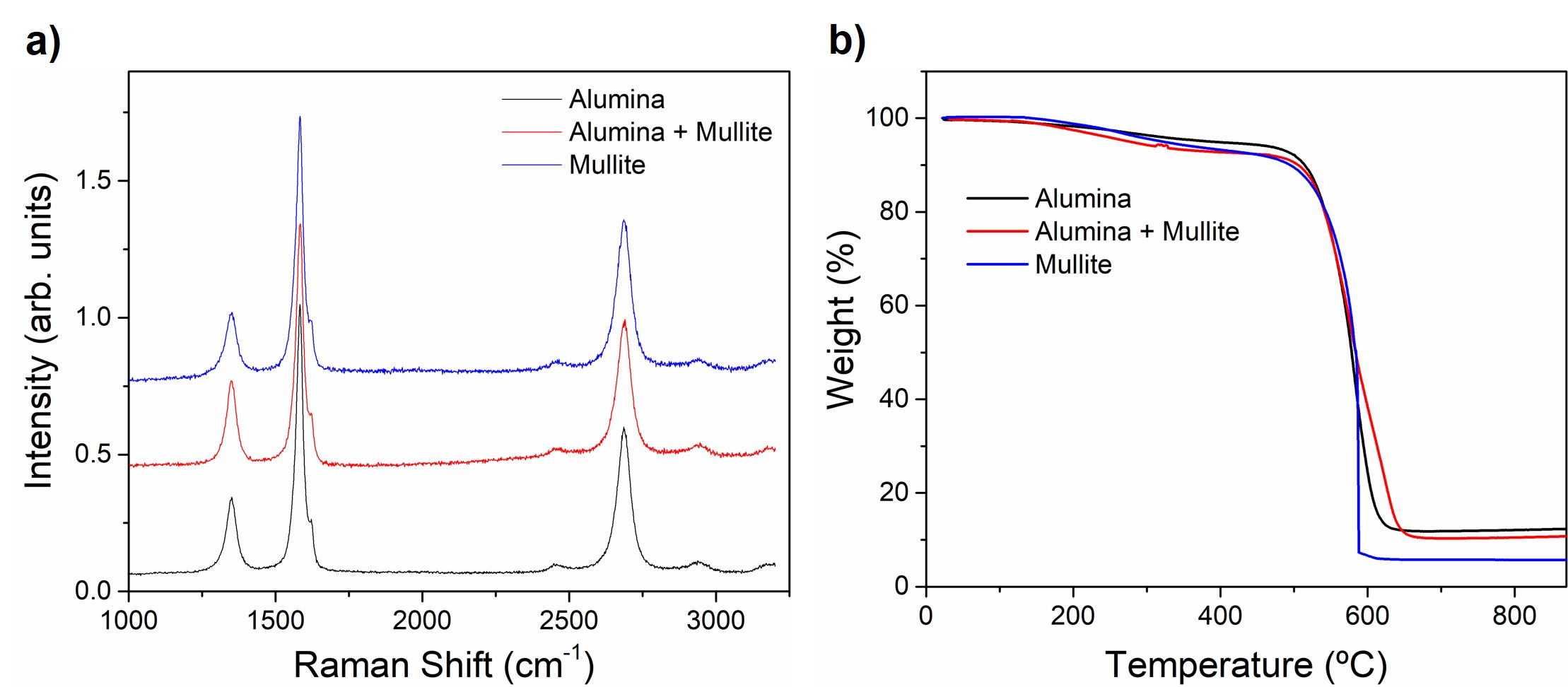}
\caption{Characterization of samples produced in different reactor tubes showing changes in reaction yield but not in CNT morphology. (a) Typical Raman spectra showing nearly identical peak positions and ratio I\textsubscript{D}/I\textsubscript{G}. (b) Thermo-gravimetric curves indicating similar composition of graphitic material and confirming higher yield for samples produced in mullite  (through a lower fraction of residual catalyst).}
\label{fig4}
\end{figure}

Interestingly, the analysis of the gases at the outlet of both reactor tubes reveals extensive thermal decomposition of the carbon source (1-butanol) irrespective of the presence of catalyst. CO$_{2}$ and C$_{1}$-C$_{3}$ hydrocarbons (methane, ethane, ethylene, acetylene) are found in all conditions studied, together with minor amounts of carbonyl sulphide. Benzene, propene and propyne are also observed in small quantities. These results are in agreement with thermal decomposition studies of n-butanol reported in the literature \cite{Harper2011}. The presence or absence of the main products does not appear to depend on the addition of catalyst or the type of reactor tube used. This result reveals that the catalyst does not play an active role in the decomposition of the carbon source, contrary to what is generally assumed. Having established that the reactor tube plays a central role in the decomposition of the C source, it is not surprising that its composition affects the concentration of different C species in the reactor. As observed in the comparison in Figure \ref{fig3_chrom}, the fraction of methane, ethylene and carbon dioxide, amongst other species, is clearly sensitive to the composition of the rector tube. Determining the precise thermodynamic pathway for the cracking of the hydrocarbons during FC-CVD synthesis of CNT fibers requires dedicated, advanced in-situ sampling methods that can operate at reaction temperatures above 1000$^{\circ}$ \cite{boies2016}, and is beyond the scope of this work. 

In spite of the substantial change in reaction conversion the type and morphology of the constituent CNTs are not significantly affected by the choice of reactor tube. The Raman spectra for samples produced in the different reactor tubes, but with the same precursor composition, presented in Figure  \ref{fig4}(a), show that in all three cases the CNTs consist of highly graphitic few-layer multi-wall nanotubes (MWNTs). The spectra reveal no appreciable change in peak positions, for instance in the G band (1583 $cm^{-1}$, 1584 $cm^{-1}$, 1583 $cm^{-1}$ for alumina, alumina with concentric mullite tube and mullite tube, respectively). Similarly, the ratio of intensities of $I\textsubscript{D}/I\textsubscript{G}$ for the three fiber types ($0.3$ for alumina, $0.4$ for alumina with mullite and $0.3$ for mullite) are within typical variation between samples and suggest no substantial differences in CNT morphology or degree of graphitization.

Thermo-gravimetric analysis (TGA) (Figure \ref{fig4}(b)) also indicates that the use of different reactor tubes has no effect on the quality and morphology of the graphitic materials, only on yield. A comparison of thermograms shows the three samples to have similar thermal decomposition curves, with nearly identical onset for CNT oxidation at 530 ºC. The content of poorly graphitized impurities, evolving at temperatures below 400 ºC, is very similar too. On the other hand, the relative content of residual catalyst also follows the trend observed in Figure \ref{fig2}a ($8.5 \%, 7.4 \% and 4.0 \%$ for mullite, mullite plus alumina and alumina, respectively), reflecting mainly an increase in graphitic material and nearly constant Fe content.  

\begin{figure}[ht]
\centering
\includegraphics[width=0.8\linewidth]{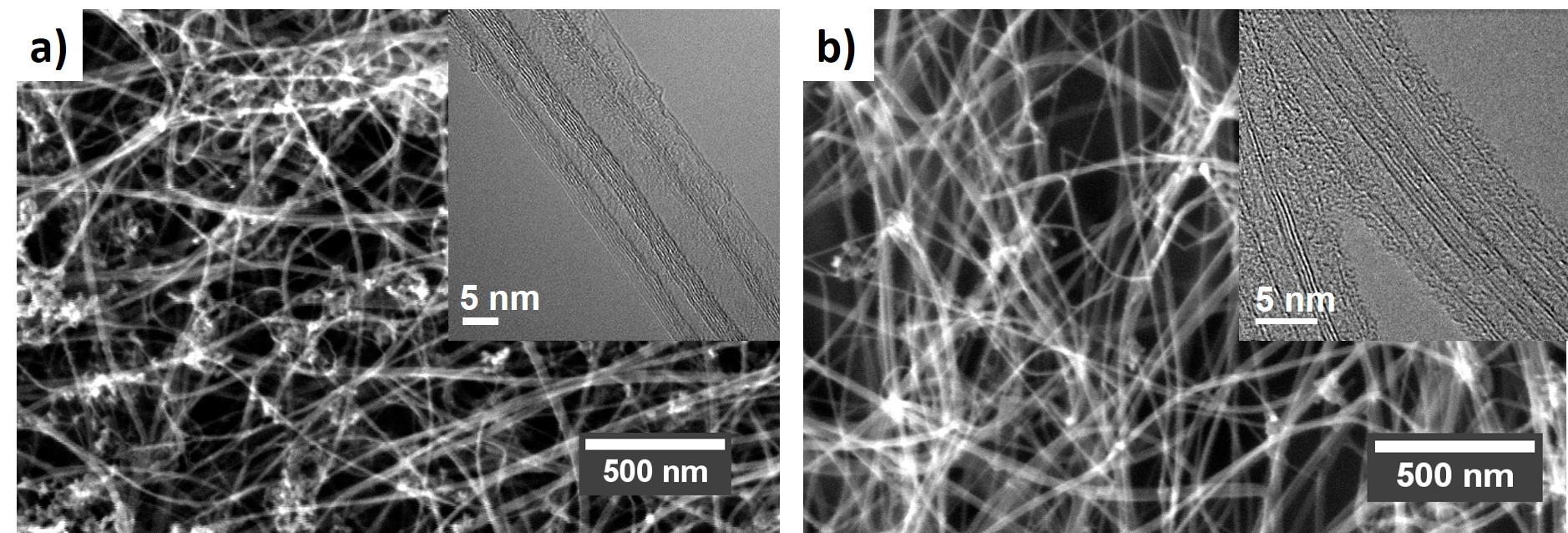}
\caption{Electron micrographs of CNT fibers produced in an alumina (a) and mullite (b) reactor tubes, showing no changes in composition or CNT type.}
\label{fig5}
\end{figure}

Electron microscopic analysis of samples produced in mullite and alumina tubes, respectively, while only being qualitative, also confirms the observations discussed above. Scanning electron micrographs (Figure \ref{fig5}(a)) gave no indication of the formation of more particulate impurities or changes in bundle morphology. Similarly, transmission electron microscopy (TEM) analysis (Figure \ref{fig5}(b)) confirmed that the CNTs produced in both reactor tubes have a high degree of graphitization (i.e. high sp$^{2}$ conjugation) and have predominantly 3-5 walls, with some of them being collapsed due to their relatively large diameter. 

\subsection*{Analysis of Si-containing impurities}
The preceding discussion leaves no doubt that the reactor tube plays an active role in the synthesis of CNTs in the direct fiber process. As a result of reactions at the reactor tube wall and of the SiO$_{2}$ in the tube itself (\textit{vide infra}), Si-containing impurities are occasionally found in CNT fibers. They are scarce, mostly found on samples collected for long times ($> 30$ min) produced in heavily used reactor tubes and when inspecting the samples with methods that probe large sample volumes and/or are sensitive to Si-containing compounds, hence why they had been undetected so far. In fact, their presence in our CNT fiber samples has been elucidated after analysis of combined data obtained using different characterization techniques over several years. Some degree of analysis of these particles is instructive in trying to understand the role of the mullite reactor tube in the CNT growth process.

\begin{figure}[ht]
\centering
\includegraphics[width=0.8\linewidth]{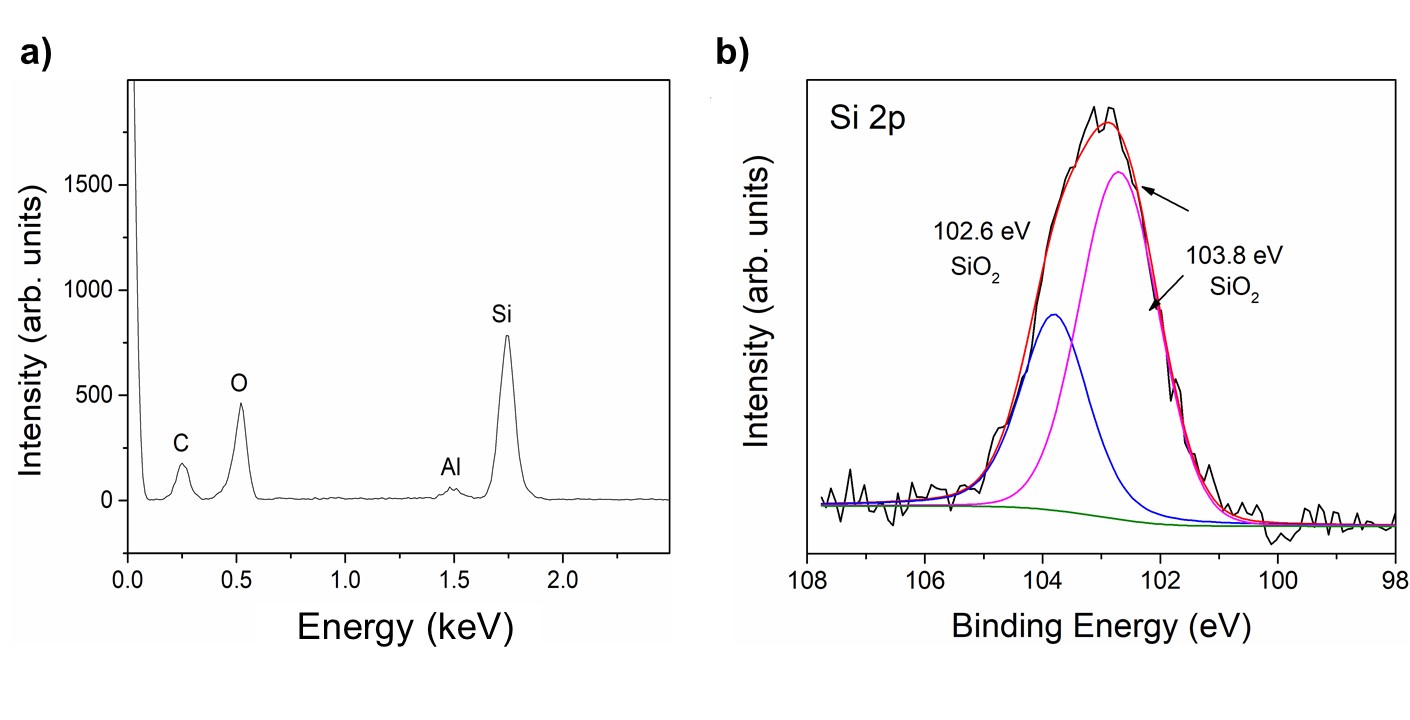}
\caption{Evidence of Si-containing impurities occasionally found in CNT fibers. (a) EDS spectrum showing a strong Si signal obtained from a thick CNT fiber sample observed under SEM. (b) Si 2p peak in the XPS spectrum of a CNT fiber, corresponding to SiO\textsubscript{2}.}
\label{fig6}
\end{figure}

Figure \ref{fig6}(a) presents an example of an energy dispersive X-ray spectroscopy (EDS) spectrum obtained from a location on a relatively thick CNT fiber sample, which gave a particularly high signal from the Si K${\alpha}$-shell emission line at 1.74 KeV. EDS intensities are strongly dependent on atomic ratio and probe a large sample volume, thus this result only confirms that Si is present in the sample.
Generally X-ray photo-electron spectroscopy (XPS) analysis of CNT fibers shows no indication of the presence Si \cite{Reguero2014,Aleman2016,BelenXPS}, however, in our extensive use of this technique we have occasionally found contributions from Si. The example in Figure \ref{fig6}(b) shows the Si 2p XPS peak, fitted by Gaussian/Lorentzian peak shapes and after Shirley background profile subtraction. The peaks at $102.6$ and $103.8$ eV are assigned to SiO\textsubscript{2},\cite{Sivakumar2011}. This observation indicates that some of the Si-containing impurities are in the form of SiO\textsubscript{2}, but without ruling out other forms of Si (e.g. SiC), particularly considering the limited ($<5$nm) probing depth of XPS.
\begin{figure}[ht]
\centering
\includegraphics[width=0.8\linewidth]{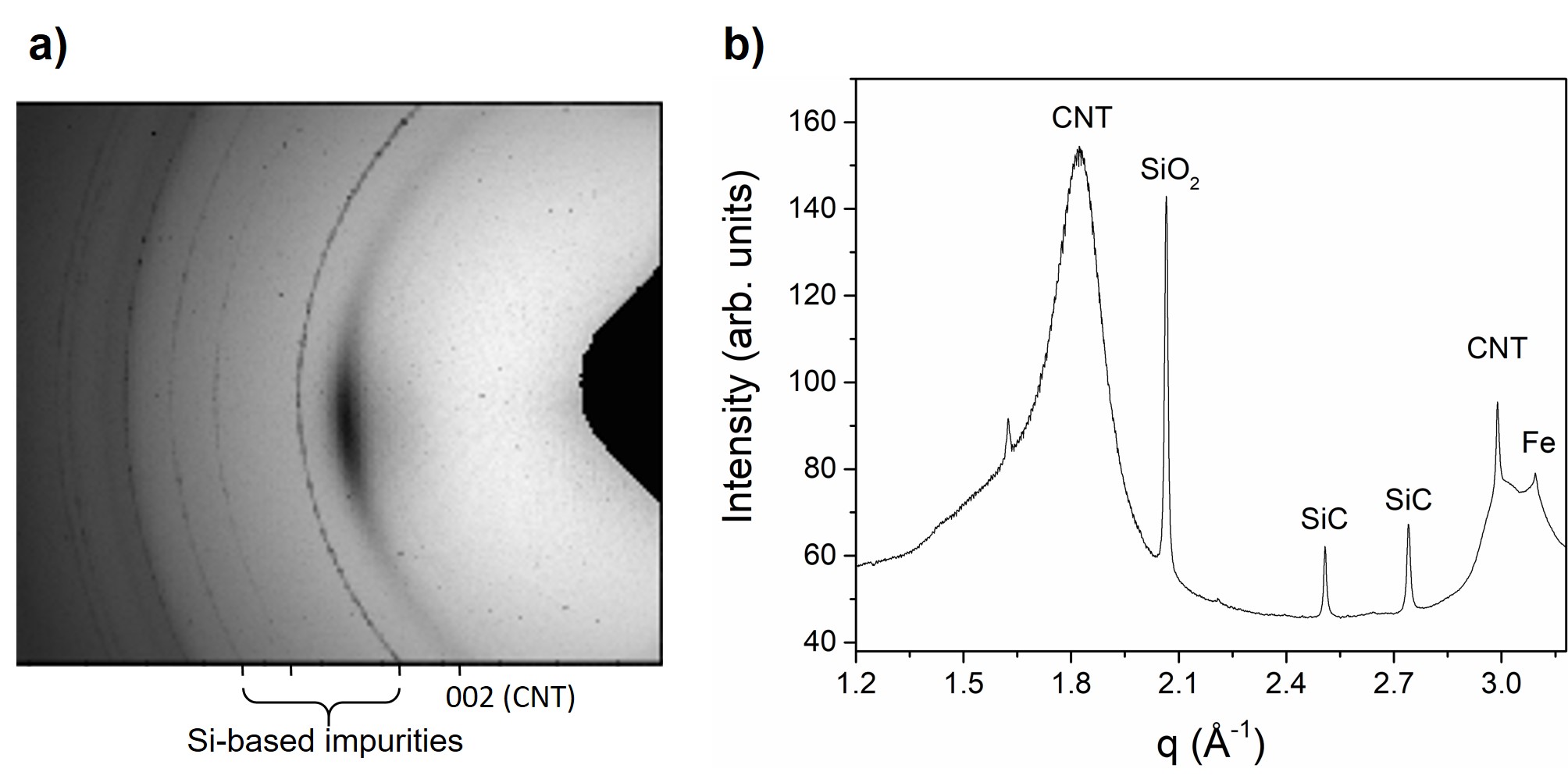}
\caption{Observation of Si-containing impurities in CNT fibers observed by synchrotron radiation WAXS in transmission mode. a) 2D WAXS pattern showing the standard reflections from CNTs and residual catalyst and the presence of sharp powder rings from impurities. b) Radial profile after integration with assignation showing reflections from SiO\textsubscript{2} and SiC. The latter implies that SiO\textsubscript{2} in the reactor tube reacts with carbon precursors.}
\label{fig7}
\end{figure}

Another evidence of Si-containing impurities in CNT fibers is found in X-ray diffraction patterns from fibers grown in mullite, our standard reactor tube. Yet, the associated features can only occasionally be resolved and only when using synchrotron radiation in transmission mode, which probes a large sample volume with a high intensity beam. An example is presented in Figure \ref{fig7}, chosen because of the particularly strong intensity of impurity reflections in this part of the sample, which favors their analysis. The same impurity features have been occasionally observed in fibers of single-wall nanotubes (SWNTs) produced with a lower S/C ratio or in samples made using different carbon precursors. 

The 2D wide-angle X-ray scattering (WAXS) pattern (Figure \ref{fig7}a) shows that in addition to the standard features of CNT fibers, namely the (002) reflection from CNTs on the equator and broad powder rings associated with the catalyst overlapped with the {100} reflection from CNTs \cite{davies2009}, there are abnormal sharp powder rings. These extraneous peaks are clearly visible in the radial profile obtained after azimuthal integration (Figure \ref{fig7}b).  Interestingly, in addition to a strong reflection at q = 2.06 $\mbox{\AA}^{-1}$ attributed to SiO\textsubscript{2} (96 – 500 – 0036) there are strong reflections at 2.50 $\mbox{\AA}^{-1}$ and 2.74 $\mbox{\AA}^{-1}$, which we assign to SiC (96 – 231 – 0936). These results do not fully determine the structure of these impurities, but they clearly show the presence of SiC, which is a key observation as it implies a reaction of carbon precursors with SiO\textsubscript{2} in the reactor tube.

Note that the Si-containing particles discussed above give rise to sharp X-ray diffraction (XRD) peaks and thus large crystal sizes ($>$ 120 nm). We take this observation as indicative that these impurities correspond to fragments of the reactor tube that detach during fiber spinning, particularly during stringing, and attach to the porous CNT fiber structure at its exit of the reactor. These fragments originate from the inside wall of the reactor tube and thus correspond to the materials that is exposed to the reaction and the highly reducing atmosphere, hence its composition is different from that of mullite.

Further analysis of these impurities is problematic because of the difficulty in locating them, since they are not associated to the catalyst or the CNTs. Conventional analysis of samples by either SEM or TEM has not revealed the presence of Si-based nanoparticles after years of studying these CNT fibers, which explains why, to our knowledge, there are no previous reports suggesting the presence of Si-based impurities after CNT growth by CVD in mullite or quartz. SiO\textsubscript{2} particles on graphene grown by CVD on Cu in hydrogen in a quartz tube have indeed been observed \cite{Benji}, but this in fact supports the view that the reactor tube degrades under nanocarbon growth conditions typically used in CVD, irrespective of the presence of a transition metal catalyst.   

In order to gain more insight into the structure of Si-containing impurities in the fibers we subjected a CNT fiber sample to ultrasonication in acetone, a known procedure to selectively remove extraneous materials from CNT fibers while preserving their CNT network structure \cite{Sundaram2017}. The resultant dispersion was deposited on a holey carbon grid to perform TEM analysis.

\begin{figure}[ht]
\centering
\includegraphics[width=0.8\linewidth]{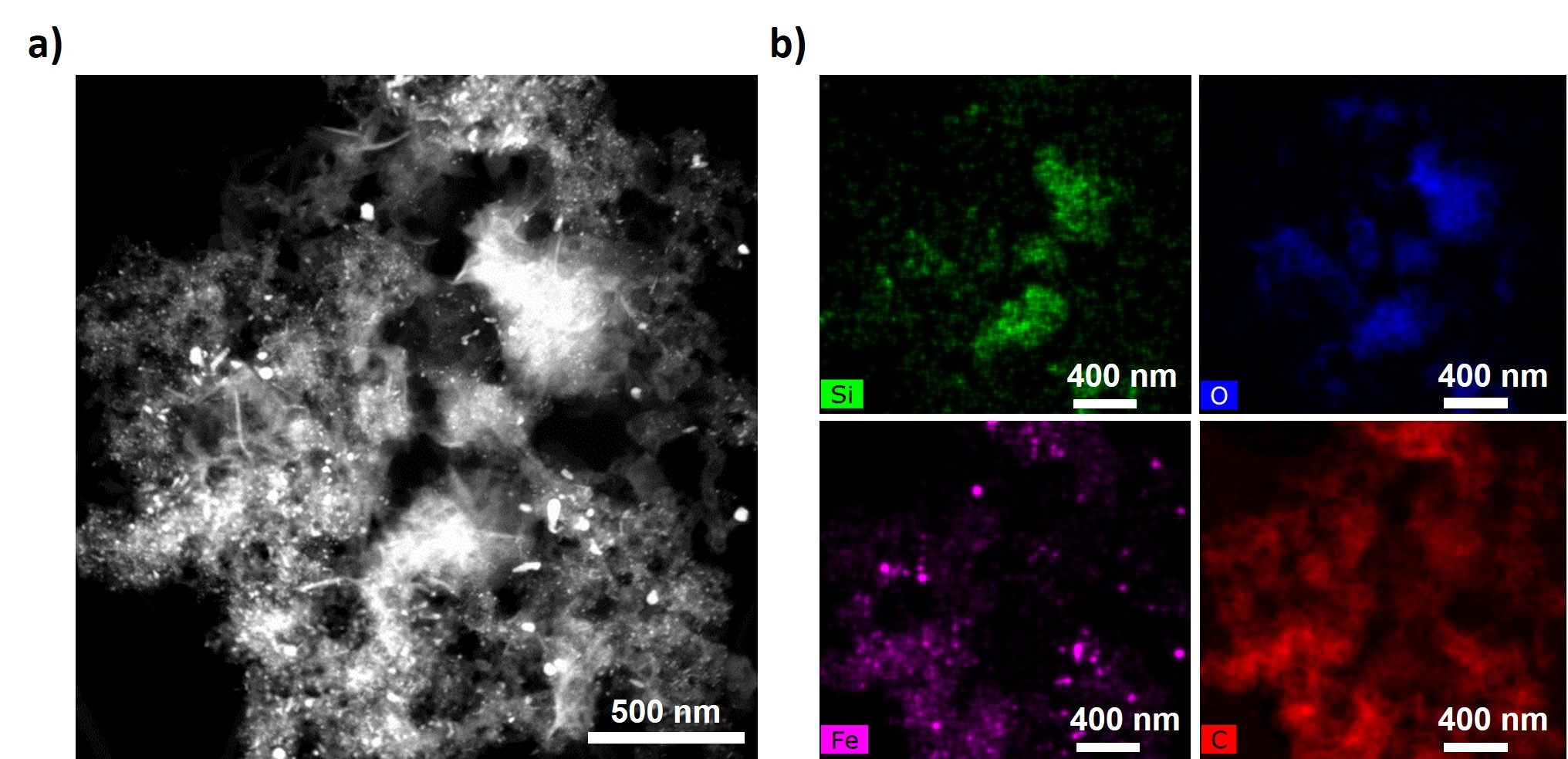}
\caption{TEM and EDS analysis of impurities obtained after dispersion. (a) TEM micrograph showing irregular particles of different sizes. (b) EDS elemental maps confirm the presence of Si, mostly associated with O but not spatially correlated with either Fe or C.}
\label{fig8}
\end{figure}

Extensive analysis of impurities showed mainly C- or Fe-based particles, as expected, but some evidence of Si-based particles was also present. Figure \ref{fig8}(a) shows an example of extracted impurities consisting of agglomerated particles of irregular morphology and size ranging from a few to hundreds of nanometers. TEM-EDS mappings performed on these agglomerated  particles clearly show that many of them are Silicon-based \ref{fig8}b. Yet, very importantly, there is no spatial correlation between the Si and Fe signals. This confirms that while playing an active role in the CNT growth process, the SiO\textsubscript{2} in the reactor tube does not directly act on the catalyst particles, nor associates with them, for example as a silicide (FeSi\textsubscript{2}).

\section*{Discussion}
The results presented above demonstrate that SiO$_{2}$ in the reactor tubes plays an active role in CVD reaction, increasing yield of CNTs but without affecting composition. They also show that some SiO$_{2}$ reacts with carbon, leading to the formation of SiC, but very importantly, there is no evidence that Si associates with the Fe-based catalyst or is even present near the catalyst in CNT fibers. While these results might be surprising in the context of CNT growth, they are fully consistent with observations in hydrocarbon heterogeneous catalysis, where SiO$_{2}$ and aluminosilicates are known to catalyze multiple reactions, including cracking of hydrocarbons. Referred to as examples of inorganic solid acids, aluminosilicates are amongst the most studied catalysts in hydrocarbon reactions, most notably, because of their widespread use in oil refining-related reactions. Extensive reviews can be found elsewhere \cite{Corma1995}, but we note some related examples that help clarify the results presented above for the high-temperature catalytic decomposition of n-butanol.      

\begin{figure}[ht]
\centering
\includegraphics[width=0.5\linewidth]{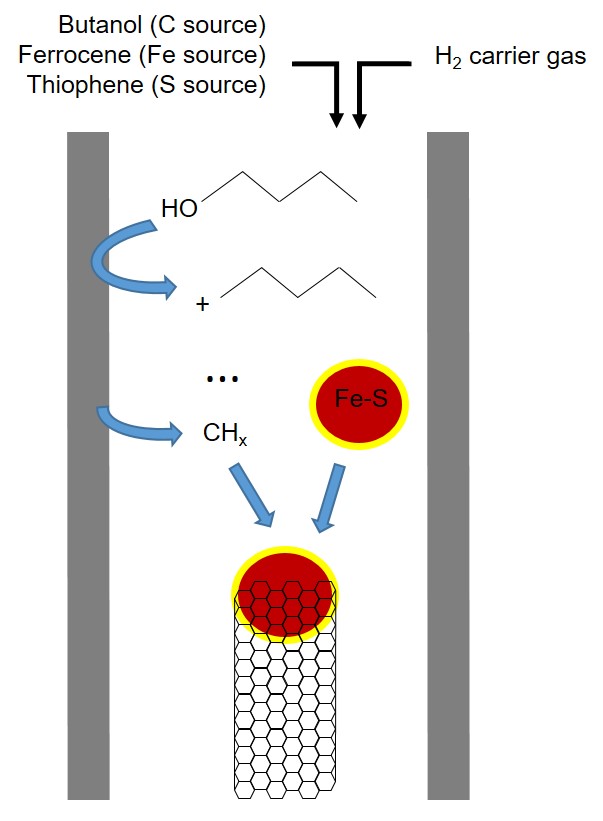}
\caption{Scheme of carbon nanotube growth in FC-CVD used for direct spinning. Carbon source (n-butanol) decomposition takes place at the reactor tube and forms smaller carbocations that finally react with the S-coated Fe catalyst particle and supply C for the nascent CNTs}
\label{fig9}
\end{figure}

The decomposition of hydrocarbons on the surface of aluminosilicates can lead to multiple path ways that produce both carbon-carbon bond rupture and formation. Early studies showed, for example, that in low-temperature (320 to 550 ºC) decomposition of a wide range of hydrocarbons (alkanes, aliphatic alcohols, etc) in the presence of an aluminosilicate with 85.6 wt.$\%$ SiO$_{2}$ and 12.9 wt. \% Al$_{2}$O$_{3}$, the rate of thermal cracking is negligible compared to the rate of catalytic cracking \cite{Earlycat}. Similarly, a monolayer of SiO$_{2}$ on Al$_{2}$O$_{3}$ has been shown to act as an electron acceptor for 2-butanol dehydration and alkane cracking, amongst other reactions \cite{Herman1966,KNOZINGER1970}. It is most likely that the SiO$_{2}$-catalyzed reaction in the direct spinning corresponds to this type, whereby n-butanol decomposes at the surface of the reactor tube and forms smaller chemical species, including carbocations, that ultimately react with the Fe-based catalyst particle and supply C for the nascent CNT, as shown in the simplified schematic in Figure \ref{fig9}. 

We note that mullite also assists CNT fiber growth when using CH$_{4}$ as a C precursor \cite{Tortech}, a common choice in scaled-up facilities. The catalytic degradation of CH$_{4}$, which is more stable than butanol, could involve the formation of its carbenium ion ($\mbox{CH$_{3}$}^{+}$), a chemical species also found in various aluminosilicate-catalyzed hydrocarbon reactions \cite{Corma1995}. In fact, examples of the effect of Si-based reactor tubes on the synthesis of graphitic materials by CVD date back to as early as 1948.\cite{Si1948}.

But to gain more insight into the extent to which the reactor tube can take part in the catalytic decomposition of C-precursors, we have calculated the approximate total surface of Fe catalyst nanoparticles relative to that of the reactor tube. Under the assumption that the residence time in the reactor is dictated by the carrier gas velocity and that the carried gas behaves as an ideal gas, we find that the ratio of surfaces can be expressed as (see SOM)
 
\begin{equation}
\label{Sratio}
\frac{S_{Fe}}{S_{reactor}} = \frac{3}{2} \frac{\dot{m}_{Fe} r_{reactor}}{r_{Fe}\rho_{Fe} f_{H_{2}}}
\end{equation}

Where $\dot{m}_{Fe}$ is the mass flow rate of Fe injected in the reactor, $r_{reactor}$ is the radius of the reactor tube ($3.5 cm$), $r_{Fe}$ the average Fe catalyst particle radius ($4.5 nm$), $\rho_{Fe}$ the density of Fe and $f_{H_{2}}$ the flow rate of the $H_{2}$ carrier gas. For typical CNT fiber spinning parameters used in laboratory settings the rate comes out as $0.05$. This be initially surprising considering that the radius of the particles is much smaller than that of the reactor tube, but note that the mass flow rate is very low (less than $3 \mu g /s$) and hence the concentration of Fe particles is extremely small, which makes the two surfaces of similar order of magnitude. Very importantly, we have observed gradual roughening of the surface of the reactor tube after its regular use over months. Such roughening occurs through repetitive use in a highly reducing atmosphere (containing H$_{2}$ and CH$_{4}$), causing partial reduction and re-oxidation (on cooling) of mullite \cite{Iso1982,Nasa1996} and its consequent amorphization (see SOM). Together with catalyst plating near the injection point at the cold end of the reactor, this could lead to even higher values of $\frac{S_{Fe}}{S_{reactor}}$ and variations in reaction yield through a higher contribution of hydrocarbon decomposition at the reactor tube.

\section*{Conclusions}
This work shows that the ceramic reactor tube used in FC-CVD synthesis of CNT fibers plays an active role in the reaction by assisting in the decomposition of the n-butanol carbon source. Using a mullite reactor leads to higher reaction yields, i.e. a higher fiber mass linear density, compared to alumina. This is consistent with their established catalytic activity for the decomposition of hydrocarbons. In spite of the key contribution of the reactor tube to the synthesis of CNT fibers, the morphology of the CNTs in terms of degree of graphitization and number of layers remains unaffected by the choice of reactor tube and is instead controlled by the ratio of promoter to carbon source. 
Si-based impurities are occasionaly observed in CNT fibers using different techniques. Their structure suggests that they correspond to reactor tube fragments that detach from the main tube and end up trapped in the porous CNT fibers. All evidence suggests that Si does not associate with the Fe catalyst nor take direct part in the extrusion of graphitic carbon at the catalyst particle surface.    
A main corollary of these results is the possibility to decouple hydrocarbon decomposition from catalytic growth, which could greatly help increase reaction yield and bring new opportunities for molecular control.This requires, amongst other advances, determining more precisely the catalytic decomposition route of different C sources and separating contributions from the Fe catalyst particles and the reactor tube to this process.

\section*{Methods}

\subsection*{Materials}
Thiophene (extra purity $\geq$ 99 $\%$) from Sigma Aldrich, ferrocene (purity $=$ 98$\%$) obtained from Acros Organics, 1-butanol and toluene (purity $>$ 99$\%$) from Sigma Aldrich were used for the synthesis of CNT fibers. Ferrocene was purified by a sublimation/recrystallization process.

\subsection*{Synthesis of CNT fibers}

Carbon nanotubes fibers were synthesized by direct spinning method. The assembly of CNTs into a fiber in this method occurs in the gas phase during growth by floating catalyst chemical vapor deposition, where the very long CNTs (1mm) entangle and form an aerogel, which is then drawn through and out of the reactor tube and continuously deposited onto a winder, \cite{Windle2004}. The reaction was carried out at 1250 ºC under hydrogen atmosphere and three kind of CNT fibers was produced in three different vertical tubular furnace reactors of alumina, alumina with mullite and mullite.
Synthesis conditions were adjusted so as to produce predominantly few-layer multiwalled CNTs (MWCNT) by using ferrocene as iron catalyst source, thiophene as sulfur catalyst promoter and butanol and toluene as carbon sources, with different relative contents ranging from (0.8 : 1.5 : 97.7) to (0.8 : 0.2 : 99.0).\cite{Reguero2014}

\subsection*{Characterization of CNT fibers}

Raman spectroscopy was performed mainly using a Raman micro-spectroscopy system Renishaw with 532 nm laser line (2.33 eV), taking three accumulations at $\approx 1.68$ mW. In order to maximize the Raman signal, the polarization of the excitation signal was kept parallel to the CNT fiber axis.

SEM micrographs and EDS spectra were taken with a FIB-FEGSEM dual-beam microscope (Helios NanoLab 600i, FEI). 
TEM samples were prepared by dispersing a small amount of CNT fiber sample in 99$\%$ pure acetone and sonicated in an ultrasonic power bath to debundle the CNTs and isolate the extraneous materials or impurities. This resulted in the formation of dark dispersion and sedimentation of material. Impurities were collected by depositing a drop of the dark dispersion on a holey carbon TEM grid.
TEM images and EDS elemental maps were taken using a FEG S/TEM microscope (Talos F200X, FEI) equipped with a chemical analysis system via EDS mode.

Thermo-gravimetric analysis were recorded under air atmosphere in a TGA Q50 (TA Instruments). The analyses were performed under air from 25ºC to 800ºC using a 10ºC/min ramp.

XPS data were collected with synchrotron radiation X-Ray source using a Phoibos 100 hemispherical energy analyzer (SPECS GmbH) in ESCAmicroscopy beamline (ELECTRA synchrotron in Trieste, Italy), using a photon energy of 650 eV. Si 2p core level spectrum was fitted by Gaussian/Lorentzian peak shapes and Shirley background profile.

2D WAXS patterns were collected at NCD BL11 ALBA Light Source using a wavelength of 1 Å. The data shown are after sample-detector distance calibration and azimuthal integration. Powder XRD measurements of the reactor tube were performed with a Bruker-AXS SMART 1000 single-crystal diffractometer using Mo KR radiation and a CCD detector. The sample detector distance was determined sing a silicon single-crystal standard. 

\subsection*{Gases sampling method}
The gas near the outlet of the reactor tube was sampled by extracting a volume of gas from inside the lower part of the reactor tube into Tedlar bags using a manual pump. The gas samples were then analyzed with a gas chromatograph (Agilent 7820 GC) coupled to a mass spectrometer (Agilent 5977B MSD). The gas was injected through an automatic valve into a CarbonPLOT column (30 m x 0.32 mm) using He as carrier gas with a split ratio of 1:20. The mass spectra were obtained by electronic impact with a single quadruple analyzer and an extractor ion source. The identification of the peaks was performed using the NIST Mass Spectral Search Program.

\bibliography{sample}

\begin{thebibliography}{10}
\expandafter\ifx\csname url\endcsname\relax
  \def\url#1{\texttt{#1}}\fi
\expandafter\ifx\csname urlprefix\endcsname\relax\def\urlprefix{URL }\fi
\expandafter\ifx\csname doiprefix\endcsname\relax\def\doiprefix{DOI }\fi
\providecommand{\bibinfo}[2]{#2}
\providecommand{\eprint}[2][]{\url{#2}}

\bibitem{senokos2016}
\bibinfo{author}{Senokos, E.}, \bibinfo{author}{Reguero, V.},
  \bibinfo{author}{Palma, J.}, \bibinfo{author}{Vilatela, J.~J.} \&
  \bibinfo{author}{Marcilla, R.}
\newblock \bibinfo{journal}{\bibinfo{title}{Macroscopic fibres of cnts as
  electrodes for multifunctional electric double layer capacitors: from quantum
  capacitance to device performance}}.
\newblock {\emph{\JournalTitle{Nanoscale}}} \textbf{\bibinfo{volume}{8}},
  \bibinfo{pages}{3620--3628} (\bibinfo{year}{2016}).

\bibitem{Neural}
\bibinfo{author}{Vitale, F.}, \bibinfo{author}{Summerson, S.~R.},
  \bibinfo{author}{Aazhang, B.}, \bibinfo{author}{Kemere, C.} \&
  \bibinfo{author}{Pasquali, M.}
\newblock \bibinfo{journal}{\bibinfo{title}{Neural stimulation and recording
  with bidirectional, soft carbon nanotube fiber microelectrodes}}.
\newblock {\emph{\JournalTitle{ACS Nano}}} \textbf{\bibinfo{volume}{9}},
  \bibinfo{pages}{4465--4474} (\bibinfo{year}{2015}).

\bibitem{Cleis}
\bibinfo{author}{Santos, C.} \emph{et~al.}
\newblock \bibinfo{journal}{\bibinfo{title}{Interconnected metal oxide cnt
  fibre hybrid networks for current collector-free asymmetric capacitive
  deionization}}.
\newblock {\emph{\JournalTitle{J. Mater. Chem. A}}} \bibinfo{pages}{--}
  (\bibinfo{year}{2018}).

\bibitem{Gonzalez2017}
\bibinfo{author}{González, C.}, \bibinfo{author}{Vilatela, J.~J.},
  \bibinfo{author}{Molina-Aldareguía, J.~M.}, \bibinfo{author}{Lopes, C.~S.}
  \& \bibinfo{author}{LLorca, J.}
\newblock \bibinfo{journal}{\bibinfo{title}{Structural composites for
  multifunctional applications: Current challenges and future trends}}.
\newblock {\emph{\JournalTitle{Progress in Materials Science}}}
  \textbf{\bibinfo{volume}{89}}, \bibinfo{pages}{194--251}
  (\bibinfo{year}{2017}).

\bibitem{Toribio2016}
\bibinfo{author}{Fernández-Toribio, J.~C.} \emph{et~al.}
\newblock \bibinfo{journal}{\bibinfo{title}{A composite fabrication sensor
  based on electrochemical doping of carbon nanotube yarns}}.
\newblock {\emph{\JournalTitle{Advanced Functional Materials}}}
  \textbf{\bibinfo{volume}{26}} (\bibinfo{year}{2016}).

\bibitem{ReviewQingwen}
\bibinfo{author}{Jiangtao, D.} \emph{et~al.}
\newblock \bibinfo{journal}{\bibinfo{title}{Carbon‐nanotube fibers for
  wearable devices and smart textiles}}.
\newblock {\emph{\JournalTitle{Advanced Materials}}}
  \textbf{\bibinfo{volume}{28}}, \bibinfo{pages}{10529--10538}
  (\bibinfo{year}{2016}).

\bibitem{Windle2004}
\bibinfo{author}{Li, Y.~L.}, \bibinfo{author}{Kinloch, I.~A.} \&
  \bibinfo{author}{Windle, A.~H.}
\newblock \bibinfo{journal}{\bibinfo{title}{Direct spinning of carbon nanotube
  fibers from chemical vapor deposition synthesis.}}
\newblock {\emph{\JournalTitle{Science}}} \textbf{\bibinfo{volume}{304}},
  \bibinfo{pages}{276--278} (\bibinfo{year}{2004}).

\bibitem{Aleman2015}
\bibinfo{author}{Alemán, B.}, \bibinfo{author}{Reguero, V.},
  \bibinfo{author}{Mas, B.} \& \bibinfo{author}{Vilatela, J.~J.}
\newblock \bibinfo{journal}{\bibinfo{title}{Strong carbon nanotube fibres by
  drawing inspiration from polymer fibre spinning.}}
\newblock {\emph{\JournalTitle{ACS Nano}}} \textbf{\bibinfo{volume}{9(7)}},
  \bibinfo{pages}{7392--7398} (\bibinfo{year}{2015}).

\bibitem{Reguero2014}
\bibinfo{author}{Reguero, V.}, \bibinfo{author}{Alemán, B.},
  \bibinfo{author}{Mas, B.} \& \bibinfo{author}{Vilatela, J.~J.}
\newblock \bibinfo{journal}{\bibinfo{title}{Controlling carbon nanotube type in
  macroscopic fibres synthesized by the direct spinning process.}}
\newblock {\emph{\JournalTitle{Chemistry of Materials}}}
  \textbf{\bibinfo{volume}{26(11)}}, \bibinfo{pages}{3550--3557}
  (\bibinfo{year}{2014}).

\bibitem{Selenio2016}
\bibinfo{author}{Mas, B.} \emph{et~al.}
\newblock \bibinfo{journal}{\bibinfo{title}{Group 16 elements control the
  synthesis of continuous fibers of carbon nanotubes.}}
\newblock {\emph{\JournalTitle{Carbon}}} \textbf{\bibinfo{volume}{101}},
  \bibinfo{pages}{458--464} (\bibinfo{year}{2016}).

\bibitem{Science}
\bibinfo{author}{Koziol, K.} \emph{et~al.}
\newblock \bibinfo{journal}{\bibinfo{title}{High-performance carbon nanotube
  fiber}}.
\newblock {\emph{\JournalTitle{Science}}} \textbf{\bibinfo{volume}{318}},
  \bibinfo{pages}{1892--1895} (\bibinfo{year}{2007}).

\bibitem{copper12017}
\bibinfo{author}{Gspann, T.} \emph{et~al.}
\newblock \bibinfo{journal}{\bibinfo{title}{High thermal conductivities of
  carbon nanotube films and micro-fibres and their dependence on morphology.}}
\newblock {\emph{\JournalTitle{Carbon}}} \textbf{\bibinfo{volume}{114}},
  \bibinfo{pages}{160--168} (\bibinfo{year}{2017}).

\bibitem{Science2013}
\bibinfo{author}{Behabtu, N.} \emph{et~al.}
\newblock \bibinfo{journal}{\bibinfo{title}{Strong, light, multifunctional
  fibers of carbon nanotubes with ultrahigh conductivity.}}
\newblock {\emph{\JournalTitle{Science}}} \textbf{\bibinfo{volume}{339}},
  \bibinfo{pages}{182--186} (\bibinfo{year}{2013}).

\bibitem{Lewaka2014}
\bibinfo{author}{Lekawa-Raus, A.}, \bibinfo{author}{Patmore, J.},
  \bibinfo{author}{Kurzepa, L.}, \bibinfo{author}{Bulmer, J.} \&
  \bibinfo{author}{Koziol, K.}
\newblock \bibinfo{journal}{\bibinfo{title}{Electrical properties of carbon
  nanotube based fibers and their future use in electrical wiring.}}
\newblock {\emph{\JournalTitle{Adv. Funct. Mater}}}
  \textbf{\bibinfo{volume}{24(4)}}, \bibinfo{pages}{3661--3682}
  (\bibinfo{year}{2014}).

\bibitem{Cantoro}
\bibinfo{author}{Cantoro, M.} \emph{et~al.}
\newblock \bibinfo{journal}{\bibinfo{title}{Catalytic chemical vapor deposition
  of single-wall carbon nanotubes at low temperatures}}.
\newblock {\emph{\JournalTitle{Nano Letters}}} \textbf{\bibinfo{volume}{6}},
  \bibinfo{pages}{1107--1112} (\bibinfo{year}{2006}).

\bibitem{Bronikowski2007}
\bibinfo{author}{Brownikowski, M.~J.}
\newblock \bibinfo{journal}{\bibinfo{title}{Longer nanotubes at lower
  temperatures: The influence of effective activation energies on carbon
  nanotube growth by thermal chemical vapor deposition}}.
\newblock {\emph{\JournalTitle{The Journal of Physical Chemistry C}}}
  \textbf{\bibinfo{volume}{111}}, \bibinfo{pages}{17705--17712}
  (\bibinfo{year}{2007}).

\bibitem{Wagner2007}
\bibinfo{author}{Fu, Q.} \& \bibinfo{author}{Wagner, T.}
\newblock \bibinfo{journal}{\bibinfo{title}{Interaction of nanostructured metal
  overlayers with oxide surfaces}}.
\newblock {\emph{\JournalTitle{Surface Science Reports}}}
  \textbf{\bibinfo{volume}{62}}, \bibinfo{pages}{431--498}
  (\bibinfo{year}{2007}).

\bibitem{Hofmann1}
\bibinfo{author}{Mattevi, C.} \emph{et~al.}
\newblock \bibinfo{journal}{\bibinfo{title}{In-situ x-ray photoelectron
  spectroscopy study of catalyst-support interactions and growth of carbon
  nanotube forests}}.
\newblock {\emph{\JournalTitle{The Journal of Physical Chemistry C}}}
  \textbf{\bibinfo{volume}{112}}, \bibinfo{pages}{12207--12213}
  (\bibinfo{year}{2008}).

\bibitem{Hofman2}
\bibinfo{author}{Hofmann, S.} \emph{et~al.}
\newblock \bibinfo{journal}{\bibinfo{title}{State of transition metal catalysts
  during carbon nanotube growth}}.
\newblock {\emph{\JournalTitle{Journal of Physical Chemistry C}}}
  \textbf{\bibinfo{volume}{113}}, \bibinfo{pages}{1648--1656}
  (\bibinfo{year}{2009}).

\bibitem{Vilatela2015}
\bibinfo{author}{Vilatela, J.~J.} \emph{et~al.}
\newblock \bibinfo{journal}{\bibinfo{title}{A spray pyrolysis method to grow
  carbon nanotubes on carbon fibres, steel and ceramic bricks}}.
\newblock {\emph{\JournalTitle{Journal of Nanoscience and Nanotechnology}}}
  \textbf{\bibinfo{volume}{15}}, \bibinfo{pages}{2858--2864}
  (\bibinfo{year}{2015}).

\bibitem{Hermite2012}
\bibinfo{author}{Delmas, M.} \emph{et~al.}
\newblock \bibinfo{journal}{\bibinfo{title}{Growth of long and aligned
  multi-walled carbon nanotubes on carbon and metal substrates}}.
\newblock {\emph{\JournalTitle{Nanotechnology}}} \textbf{\bibinfo{volume}{23}},
  \bibinfo{pages}{105604} (\bibinfo{year}{2012}).

\bibitem{moisala2003}
\bibinfo{author}{Moisala, A.}, \bibinfo{author}{Nasibulin, A.} \&
  \bibinfo{author}{I.~Kauppinen, E.}
\newblock \bibinfo{journal}{\bibinfo{title}{The role of metal nanoparticles in
  the catalytic production of single-walled carbon nanotubes — a review}}.
\newblock {\emph{\JournalTitle{J. Phys.: Condens. Matter}}}
  \textbf{\bibinfo{volume}{15}}, \bibinfo{pages}{3011--3035}
  (\bibinfo{year}{2003}).

\bibitem{moisala2005}
\bibinfo{author}{Moisala, A.}, \bibinfo{author}{Nasibulin, A.},
  \bibinfo{author}{D.~Shandakov, S.}, \bibinfo{author}{Jiang, H.} \&
  \bibinfo{author}{I.~Kauppinen, E.}
\newblock \bibinfo{journal}{\bibinfo{title}{On-line detection of single-walled
  carbon nanotube formation during aerosol synthesis methods}}.
\newblock {\emph{\JournalTitle{Carbon}}} \textbf{\bibinfo{volume}{43}},
  \bibinfo{pages}{2066--2074} (\bibinfo{year}{2005}).

\bibitem{moisala2006}
\bibinfo{author}{Moisala, A.} \emph{et~al.}
\newblock \bibinfo{journal}{\bibinfo{title}{Single-walled carbon nanotube
  synthesis using ferrocene and iron pentacarbonyl in a laminar flow reactor}}.
\newblock {\emph{\JournalTitle{Chemical Engineering Science}}}
  \textbf{\bibinfo{volume}{61}}, \bibinfo{pages}{4393--4402}
  (\bibinfo{year}{2006}).

\bibitem{Sundaram2011}
\bibinfo{author}{Sundaram, R.~M.}, \bibinfo{author}{Koziol, K.~K.~K.} \&
  \bibinfo{author}{Windle, A.~H.}
\newblock \bibinfo{journal}{\bibinfo{title}{Continuous direct spinning of
  fibers of single-walled carbon nanotubes with metallic chirality.}}
\newblock {\emph{\JournalTitle{Advances Materials}}}
  \textbf{\bibinfo{volume}{23}}, \bibinfo{pages}{5064--5068}
  (\bibinfo{year}{2011}).

\bibitem{Harper2011}
\bibinfo{author}{Harper, M.~R.}, \bibinfo{author}{Geem, K. M.~V.},
  \bibinfo{author}{Pyl, S.~P.}, \bibinfo{author}{Marin, G.~B.} \&
  \bibinfo{author}{Green, W.~H.}
\newblock \bibinfo{journal}{\bibinfo{title}{Comprehensive reaction mechanism
  for n-butanol pyrolysis and combustion}}.
\newblock {\emph{\JournalTitle{Combustion and Flame}}}
  \textbf{\bibinfo{volume}{158}}, \bibinfo{pages}{16--41}
  (\bibinfo{year}{2011}).

\bibitem{boies2016}
\bibinfo{author}{Hoecker, C.}, \bibinfo{author}{Smail, F.},
  \bibinfo{author}{Pick, M.} \& \bibinfo{author}{Boies, A.}
\newblock \bibinfo{journal}{\bibinfo{title}{The influence of carbon source and
  catalyst nanoparticles on cvd synthesis of cnt aerogel}}.
\newblock {\emph{\JournalTitle{Chemical Engineering Journal}}}
  \textbf{\bibinfo{volume}{314}} (\bibinfo{year}{2016}).

\bibitem{Aleman2016}
\bibinfo{author}{Aleman, B.} \emph{et~al.}
\newblock \bibinfo{journal}{\bibinfo{title}{Inherent predominance of high
  chiral angle metallic carbon nanotubes in continuous fibers grown from molten
  catalyst}}.
\newblock {\emph{\JournalTitle{Nanoscale}}} \textbf{\bibinfo{volume}{8}},
  \bibinfo{pages}{4236--4244} (\bibinfo{year}{2016}).

\bibitem{BelenXPS}
\bibinfo{author}{Belén, A.}, \bibinfo{author}{María, V.} \&
  \bibinfo{author}{Juan~J., V.}
\newblock \bibinfo{journal}{\bibinfo{title}{Surface chemistry analysis of
  carbon nanotube fibers by x-ray photoelectron spectroscopy}}.
\newblock {\emph{\JournalTitle{Phys. Status Solidi A}}}
  (\bibinfo{year}{2018}).

\bibitem{Sivakumar2011}
\bibinfo{author}{Sivakumar, M.}, \bibinfo{author}{Venkatakrishnan, K.} \&
  \bibinfo{author}{Tan, B.}
\newblock \bibinfo{journal}{\bibinfo{title}{Characterization of mhz pulse
  repetition rate femtosecond laser-irradiated gold-coated silicon surfaces.}}
\newblock {\emph{\JournalTitle{Nanoscale Res Lett}}}
  \textbf{\bibinfo{volume}{6: 78}} (\bibinfo{year}{2011}).

\bibitem{davies2009}
\bibinfo{author}{Davies, J.~R.}, \bibinfo{author}{Riekel, C.},
  \bibinfo{author}{Koziol, K.}, \bibinfo{author}{J.~Vilatela, J.~J.} \&
  \bibinfo{author}{Windle, A.}
\newblock \bibinfo{journal}{\bibinfo{title}{Structural studies on carbon
  nanotube fibres by synchrotron radiation microdiffraction and
  microfluorescence}}.
\newblock {\emph{\JournalTitle{Journal of Applied Crystallography}}}
  \textbf{\bibinfo{volume}{42}}, \bibinfo{pages}{1122--1128}
  (\bibinfo{year}{2009}).

\bibitem{Benji}
\bibinfo{author}{Islam, A.~E.} \emph{et~al.}
\newblock \bibinfo{journal}{\bibinfo{title}{Atomic level cleaning of
  poly-methyl-methacrylate residues from the graphene surface using radiolized
  water at high temperatures}}.
\newblock {\emph{\JournalTitle{Applied Physics Letters}}}
  \textbf{\bibinfo{volume}{111}}, \bibinfo{pages}{103101}
  (\bibinfo{year}{2017}).

\bibitem{Sundaram2017}
\bibinfo{author}{M.~Sundaram, R.} \& \bibinfo{author}{Windle, A.}
\newblock \bibinfo{journal}{\bibinfo{title}{One-step purification of
  direct-spun cnt fibers by post-production sonication}}.
\newblock {\emph{\JournalTitle{Materials $\&$ Design}}}
  \textbf{\bibinfo{volume}{126}}, \bibinfo{pages}{85 -- 90}
  (\bibinfo{year}{2017}).

\bibitem{Corma1995}
\bibinfo{author}{Corma, A.}
\newblock \bibinfo{journal}{\bibinfo{title}{Inorganic solid acids and their use
  in acid-catalyzed hydrocarbon reactions}}.
\newblock {\emph{\JournalTitle{Chemical Reviews}}}
  \textbf{\bibinfo{volume}{95}}, \bibinfo{pages}{559--614}
  (\bibinfo{year}{1995}).

\bibitem{Earlycat}
\bibinfo{author}{Franklin, J.~L.} \& \bibinfo{author}{Nicholson, D.~E.}
\newblock \bibinfo{journal}{\bibinfo{title}{A kinetic study of the
  decomposition of hydrocarbons by silica–alumina catalysts}}.
\newblock {\emph{\JournalTitle{The Journal of Physical Chemistry}}}
  \textbf{\bibinfo{volume}{60}}, \bibinfo{pages}{59--62}
  (\bibinfo{year}{1956}).

\bibitem{Herman1966}
\bibinfo{author}{Pines, H.} \& \bibinfo{author}{Manassen, J.}
\newblock \bibinfo{journal}{\bibinfo{title}{The mechanism of dehydration of
  alcohols over alumina catalysts}}.
\newblock {\emph{\JournalTitle{Advan. Catal. rel. Subj.}}}
  \textbf{\bibinfo{volume}{16}}, \bibinfo{pages}{49--93}
  (\bibinfo{year}{1966}).

\bibitem{KNOZINGER1970}
\bibinfo{author}{Knözinger, H.} \& \bibinfo{author}{Scheglila, A.}
\newblock \bibinfo{journal}{\bibinfo{title}{The dehydration of alcohols on
  alumina: Xii. kinetic isotope effects in the olefin formation from
  butanols}}.
\newblock {\emph{\JournalTitle{Journal of Catalysis}}}
  \textbf{\bibinfo{volume}{17}}, \bibinfo{pages}{252--263}
  (\bibinfo{year}{1970}).

\bibitem{Tortech}
\bibinfo{author}{Winer, I.}
\newblock \bibinfo{howpublished}{personal communication}.

\bibitem{Si1948}
\bibinfo{author}{Iley, R.} \& \bibinfo{author}{Riley, H.}
\newblock \bibinfo{journal}{\bibinfo{title}{The deposition of carbon on
  vitreous silicas}}.
\newblock {\emph{\JournalTitle{Journal of the Chemical Society}}}
  \textbf{\bibinfo{volume}{0}}, \bibinfo{pages}{1362--1366}
  (\bibinfo{year}{1948}).

\bibitem{Iso1982}
\bibinfo{author}{Iso, S.~T.} \& \bibinfo{author}{Pask, J.~A.}
\newblock \bibinfo{journal}{\bibinfo{title}{Reaction of silicate glasses and
  mullite with hydrogen gas}}.
\newblock {\emph{\JournalTitle{Journal of the American Ceramic Society}}}
  \textbf{\bibinfo{volume}{65}}, \bibinfo{pages}{383--387}
  (\bibinfo{year}{1982}).

\bibitem{Nasa1996}
\bibinfo{author}{Herbell, T.~P.}, \bibinfo{author}{Hull, D.~R.} \&
  \bibinfo{author}{Garg, A.}
\newblock \bibinfo{journal}{\bibinfo{title}{Hot hydrogen exposure degradation
  of the strength of mullite}}.
\newblock {\emph{\JournalTitle{Journal of the American Ceramic Society}}}
  \textbf{\bibinfo{volume}{81}} (\bibinfo{year}{1996}).

\end{thebibliography}

\section*{Acknowledgements}

The authors are grateful for generous financial support provided by the European Union Seventh Framework Program under grant agreements 678565 (ERC-STEM), the Madrid Regional Government (MAD2D, S2013/MIT-3007) and the Spanish Ministry of Science, Innovation and Universities (Project FOTOFUEL: ENE2016-82025-REDT). F.F. and V.A.P.O. thank financial support from the European Research Council (HyMAP programme, 648319) and the Spanish Ministry of Science, Innovation and Universities (Project SOLPAC: ENE2017-89170-R, MCIU/AEI/FEDER). M.V. acknowledges the Madrid Regional Government (program “Atracción de Talento Investigador”, 2017-T2/IND-5568) for financial support. Support from NCD beam-line staff at ALBA Synchrotron Light Facility for assistance with synchrotron experiments and from ESCA microscopy beam-line staff (Elettra synchrotron, Italy) for SPEM experiments is also acknowledged.

\section*{Author contributions statement}

J.J.V, V.P. and F.F. conceived the experiments,  X.R., V.R., M.V., B.A., F.F. and L.A. conducted the experiments. All authors analyzed the results and reviewed the manuscript. 

\section*{Additional information}

The authors declare no competing interests. 

\end{document}